\documentclass[useAMS,referee]{mn2e}

\usepackage{epsfig,rotate,graphicx}

\title{Protoplanets with core  masses below the critical mass fill 
in their Roche lobe}

\author[Terquem and Heinemann]{Caroline Terquem$^{1,3}$\thanks{E-mail:
caroline.terquem@iap.fr; tobi@ias.edu} 
and Tobias Heinemann$^2$\footnotemark[1] \\ $^1$ Institut
d'Astrophysique de Paris, UPMC Univ Paris 06, CNRS, UMR7095, 98 bis bd
Arago, 75014, Paris, France \\ 
$^2$ Institute for Advanced Study, Einstein Drive, Princeton, NJ 08540, USA\\
$^3$ Princeton University, Peyton Hall, 4 Ivy Lane, Princeton, NJ 08540, USA }

\begin{document}

\maketitle

\begin{abstract}
We study the evolution of a protoplanet of a few earth masses embedded
in a protoplanetary disc.  If we assume that the atmosphere of the
protoplanet, i.e. the volume of gas in hydrostatic equilibrium bound
to the core, has a surface radius smaller than the Roche lobe radius,
we show that it expands as it accretes both planetesimals and gas at a
fixed rate from the nebula until it fills in the Roche lobe.  The
evolution occurs on a timescale shorter than the formation or
migration timescales.  Therefore, we conclude that protoplanets of a
few earth masses have an atmosphere that extends to the Roche lobe
surface, where it matches to the nebula.  This is true even when the
Bondi radius is smaller than the Roche lobe radius.  This is in
contrast to the commonly used models in which the static atmosphere
extends up to the Bondi radius and is surrounded by a cold accretion
flow.  As a result, any calculation of the tidal torque exerted by the
disc onto the protoplanet should exclude the material present in the
Roche lobe, since it is bound to the protoplanet.
 
\end{abstract}


\section{Introduction}
\label{sec:intro}

The recent release of Kepler data reporting more than 1200 planet
candidates transiting their host star confirms that the core accretion
scenario for forming planets is ubiquitous.  Indeed, most of these
objects are less than half the radius of Jupiter, which indicates that
they are at most Neptune--size (Borucki et al. 2011).  Such low--mass
objects are probably not the result of gravitational instabilities,
but are more likely formed through the accumulation of a metal rich
core followed by the capture of a more or less massive envelope of gas
(Lissauer 1993 and references therein).

In this model, the core is believed to be formed through the
solid--body accretion of kilometre--sized planetesimals.  For typical
disc parameters, when the core reaches about 0.1 earth mass, it starts
binding the gas of the nebula in which it is embedded and a gaseous
atmosphere forms around it.  As long as the core is less massive than
the so--called {\em critical mass}, the energy radiated from the
atmosphere is compensated for by the gravitational energy that the
planetesimals entering the atmosphere release when they collide with
the surface of the core.  During this stage of evolution, the
atmosphere is therefore in quasi--static and thermal equilibrium and
grows slowly in mass along with the core (Perri \& Cameron 1974,
Mizuno 1980).  By the time the core reaches the critical mass, the
atmosphere has become too massive to be supported at equilibrium by
the energy released by the planetesimals.  At that point, it starts
contracting and the subsequent runaway accretion of gas leads to the
formation of giant planets (Bodenheimer \& Pollack 1986, Pollack et
al. 1996).

At the same time as they form, protoplanets migrate through the nebula
as a result of tidal interaction with the surrounding gas (Goldreich
and Tremaine 1979, 1980, Lin and Papaloizou 1979, 1993 and references
therein, Papaloizou \& Lin 1984, Ward 1986, 1997).  Calculations
performed in isothermal discs show that cores of several earth masses
migrate inward on a relatively short timescale, shorter than the
planet formation timescale (Ward 1997, Tanaka et al. 2002, Bate et
al. 2003).  However, more recent calculations including detailed
energy balance in non isothermal discs show that migration can be
slower or even outward (Paardekooper \& Mellema 2006, Kley \& Crida
2008, Paardekooper \& Mellema 2008, Baruteau \& Masset 2008,
Paardekooper \& Papaloizou 2008).  Usually, the torque exerted by the
disc on the protoplanet is calculated by excluding the gas comprised
in its Roche lobe.  However, it has been suggested by D'Angelo et
al. (2003, see also Crida et al. 2009) that the gas present in the
Roche lobe but not bound to the protoplanet may contribute
significantly to the total torque and slow down migration.  This
happens when the atmosphere of the protoplanet, i.e. the volume of gas
at hydrostatic equilibrium that is bound to the core, does not fill in
its Roche lobe, and is surrounded by a cold (Bondi type) accretion flow.

Models of atmospheres of protoplanets that do not fill in the Roche
lobe have been considered by different authors.  Papaloizou \& Nelson
(2005) have constructed models assuming that the only source of energy
for the atmosphere is that produced by the gravitational contraction
of the gas.  This corresponds to the runaway gas accretion phase,
i.e. after the core has reached the critical mass, as the luminosity
due to the accretion of planetesimals is then negligible.  The radius
of the protoplanet atmosphere may become smaller than the Roche lobe
radius in this phase if the protostellar disc cannot supply gas to the
atmosphere rapidly enough.  Such models have also been considered by
Mordasini et al. (2011).  Lissauer et al. (2009, see also Movshovitz
et al. 2010) have calculated models of atmospheres prior to the
runaway gas accretion phase by assuming that the surface radius is
either the Bondi radius or a fraction of the Roche lobe radius
depending on whether the Bondi radius is smaller or larger than the
Roche lobe radius.  This is based on the results of three dimensional
numerical simulations indicating that only gas within about one
quarter of the Roche lobe remains bound to the protoplanet.  The gas
accretion rate onto the atmosphere is then taken to be the value
required to match the outer radius to the desired value.  To calculate
the volume of gas which is bound to the protoplanet from the numerical
simulations, they identify the trajectories of tracer particles that
are trapped inside the gravitational potential of the protoplanet.  We
comment that this approach may severely underestimate the amount of
gas that can be accreted as in reality particles can collide with the
protoplanet and become bound if energy is dissipated into shocks, as
in the process of star formation.  If the protoplanet fills in a
significant part of its Roche lobe, collisions may become frequent.

Three dimensional numerical simulations of protoplanets of a few earth
masses in protoplanetary discs usually start with a core surrounded by
a static atmosphere that extends at most up to the Bondi radius.  The
Keplerian flow located within and around the Roche lobe, and in which
the protoplanet is embedded, is then perturbed and material from the
disc rains down on the protoplanet from above and below, because the
gravitational potential from the protoplanet is not balanced (D'Angelo
et al. 2003, Paardekooper \& Mellema 2008, Machida et al. 2010).  This
results in the protoplanet being surrounded by a cold isothermal
accretion flow.  In the present paper, we show that, if a core below
the critical mass is surrounded by an atmosphere that does not fill in
entirely the Roche lobe and is embedded in an accretion flow, it is
going to evolve into a static model expanded to the Roche lobe radius
on a very short timescale, even when the Bondi radius is smaller than
the Roche lobe radius.

In the calculations presented here, we include the accretion of
planetesimals onto the core, as we focus on the stage of the evolution
before runaway gas accretion starts.  In section~\ref{sec:coremass},
we calculate the mass of the atmosphere that can be supported at
hydrostatic and thermal equilibrium around a core of a given mass
assuming the surface radius of the atmosphere to be fixed.  We
consider a range of surface radii from a few core radii up to the
Roche lobe radius.  In section~\ref{sec:evolution}, we consider the
evolution of a core that starts accreting gas at a given rate from a
radius much smaller than its Roche lobe radius.  In contrast to what
is usually assumed, we suppose here that the gas accretion rate onto
the protoplanet is fixed, as it is likely to be constrained by the
disc's physical parameters (e.g., density, temperature).  We calculate
a sequence of quasi--static equilibria along which the mass of the
protoplanet increases.  We show that the atmosphere has to expand very
rapidly to accommodate the gas that is being accreted, even if the
accretion rate is relatively small.  We argue that, even when the
Bondi radius is smaller than the Roche lobe radius, the protoplanet
will eventually fill in its Roche lobe.
Finally, in section~{\ref{sec:discussion}, we summarise
and discuss our results.

\section{Dependence of the atmosphere mass on the atmosphere surface radius}
\label{sec:coremass}

We consider an atmosphere at hydrostatic and thermal equilibrium and
calculate its mass for a fixed core mass, accretion rate of
planetesimals and surface radius of the atmosphere.

\subsection{Structure of the protoplanet atmosphere}
\label{sec:envelope}

The equations governing the structure of the protoplanet atmosphere at
hydrostatic and thermal equilibrium are the same as the equations of
stellar structure, with the nuclear energy rate being replaced by the
rate at which planetesimals that enter the planet atmosphere and collide
with the core surface release their energy.  These equations have been
presented in Papaloizou \& Terquem (1999) and we recall them below.

We assume that the protoplanet is spherically symmetric and
nonrotating.  We denote $r$ the radius in spherical coordinates in a
frame with origin at the centre of the protoplanet.  The equation of
hydrostatic equilibrium is:

\begin{equation}
\frac{dP}{d r} = - g \rho .
\label{dPdr}
\end{equation}

\noindent Here, $P$ is the pressure, $g=G M(r) / r^2$ is the
acceleration due to gravity, with $M(r)$ being the mass interior to
radius $r$ (this includes the core mass if $r$ is larger than the core
radius) and $G$ is the gravitational constant.  The relation between
$M(r)$ and the mass density per unit volume $\rho$ is:

\begin{equation}
\frac{dM}{d r} = 4 \pi r^2 \rho.
\label{dMdr}
\end{equation}

We adopt the equation of state for a hydrogen and helium mixture given
by Chabrier et al. (1992) for mass fractions of hydrogen and helium of
0.7 and 0.28, respectively.  The standard equation of radiative
transport gives the luminosity $L_{\rm rad}$ that is transported by
radiation through the atmosphere as a function of the temperature
gradient $dT/dr$:

\begin{equation}
\frac{dT}{d r} = \frac{-3 \kappa \rho}{16 \sigma
T^3} \frac{L_{\rm rad}}{4 \pi r^2} ,
\label{dTdr}
\end{equation}

\noindent where $\kappa$ is the opacity, which in
general depends on both $\rho$ and $T$, and $\sigma$ is the
Stefan--Boltzmann constant.

As a large part of the atmosphere interior is unstable to convection,
only a fraction of the total luminosity is transported by radiation.
In the models presented here, we assume that the only energy source
comes from the planetesimals that are accreted by the protoplanet and
release their gravitational energy as they collide with the surface of
the core.  They generate a total core luminosity $L_{\rm core}$ given by:

\begin{equation}
L_{\rm core} = \frac{ G M_{c} \dot{M}_{c}}{ r_{c}} ,
\end{equation}

\noindent where $M_{c}$ and $r_{c}$ are, respectively, the mass and
the radius of the core, and $\dot{M}_{c}$ is the planetesimal
accretion rate.  

The radiative and adiabatic temperature gradients,
$\nabla_{\rm rad}$ and $\nabla_{\rm ad}$, are given by:

\begin{equation}
\nabla_{\rm rad} = \left( \frac{\partial \ln T}{\partial \ln P}
\right)_{\rm rad} = \frac{3 \kappa L_{\rm core} P}{64 \pi
\sigma G M T^4} ,
\label{dTdr_rad}
\end{equation}

\noindent and

\begin{equation}
\nabla_{\rm ad} = \left( \frac{\partial \ln T}{\partial \ln P} \right)_{s} ,
\end{equation}

\noindent with the subscript ${s}$ denoting evaluation at constant
entropy.

\noindent In the regions where $\nabla_{\rm rad} < \nabla_{\rm ad}$,
there is stability to convection and therefore all the energy is
transported by radiation, i.e. $L_{\rm rad}=L_{\rm core}$.  In the regions
where $\nabla_{\rm rad} > \nabla_{\rm ad}$, there is instability to
convection and therefore part of the energy is transported by
convection, i.e.  $L_{\rm core}= L_{\rm rad}+L_{\rm conv}$, where
$L_{\rm conv}$ is the luminosity associated with convection.  
Using the
mixing length theory (Cox~\& Giuli 1968), we obtain:


\begin{equation}
L_{\rm conv}  =  \pi r^2 C_p \Lambda_{\rm ml}^2 \left[ \left( \frac{\partial
T}{\partial r} \right)_{s} - \left( \frac{\partial T}{\partial r}
\right) \right]^{3/2}
\sqrt{ \frac{1}{2} \rho g \left| \left(
\frac{\partial \rho}{\partial T} \right)_P \right| } ,
\end{equation}

\noindent where $\Lambda_{\rm ml}=|\alpha_{ml}P/(dP/d r)|$ is the
mixing length, $\alpha_{ml}$ being a constant of order unity, $\left(
\partial T/\partial r \right)_{s} = \nabla_{\rm ad} T \left( d \ln
P / d r \right)$, and the subscript $P$ denotes evaluation at constant
pressure.  The different thermodynamic parameters needed in the above
equation are given by Chabrier et al. (1992), and we fix
$\alpha_{\rm ml}=1$.

\subsection{Boundary conditions}
\label{sec:boundary}

We have to solve the above equations for the three variables $P$, $M$
and $T$ as a function of $r$.  Accordingly, we need three boundary
conditions.  

Following Bodenheimer~\& Pollack (1986) and Pollack et al. (1996), we take
for the mass density of the core $\rho_{c} = 3.2$~g~cm$^{-3}$.  The
inner boundary of the atmosphere is equal to the core radius $r_c$,
given by:
\begin{equation}
r_{c} = \left( \frac{3 M_{c}}{4 \pi \rho_{c}} \right)^{1/3}.
\end{equation}
The first boundary condition is that $M(r_c)=M_{c}.$

We denote $r_{\rm atm}$ the outer boundary of the atmosphere, and we
assume this radius to be fixed.  In the calculations presented below,
we will take $r_{\rm atm}$ to be either a few core radii or the Roche
lobe radius $r_L$, which is assumed to be the same as the Hill radius,
and is given by:

\begin{equation}
r_L= \frac{2}{3} \left( \frac{M \left( r_{\rm atm}
\right)}{3 M_{\star}} \right)^{1/3} a, 
\end{equation}

\noindent where $a$ is the orbital radius of the protoplanet and
$M_{\star}$ is the mass of the central star.  Throughout the paper, we
take $M_{\star}=1~{\rm M}_{\odot}.$

When $r_{\rm atm} < r_L$, we assume that the space above the
atmosphere and in the Roche lobe of the protoplanet is filled with a
cold accretion flow.  Then the temperature $T_s$ and the pressure
$P_s$ at the surface $r=r_{\rm atm}$ can be approximated by the
standard photospheric conditions (e.g., Kippenhahn \& Weigert 1990,
Bodenheimer et al. 2000):

\begin{equation}
T_s = \left( \frac{ L_{\rm core}}{4 \pi \sigma r_{\rm atm}^2}
\right)^{1/4},
\label{eqTs}
\end{equation}

\noindent where $\sigma$ is the Stefan--Boltzmann constant, and:

\begin{equation}
P_s = \frac{GM \left( r_{\rm atm} \right)}{r_{\rm atm}^2}
 \frac{2}{3 \kappa_s},
\end{equation}

\noindent where $\kappa_s = \kappa \left( \rho_s, T_s \right)$ and
$\rho_s$ is the mass density at $r=r_{\rm atm}$.  Similar boundary
conditions have been used by Palla \& Stahler (1992) for computing the
evolution of a protostar accreting from a disc.  The validity of these
boundary conditions will be discussed in section~\ref{sec:accretion}.
Note that in any case the structure of the atmosphere does not depend
significantly on $T_s$ (Papaloizou \& Terquem 1999).


We also present below some calculations with $r_{\rm atm}=r_L$, which are
done using the boundary conditions presented in Papaloizou \& Terquem
(1999).

\subsection{Calculations}
\label{sec:envelope_calc}

For a given core mass $M_c$, planetesimal accretion rate $\dot{M}_c $
and atmosphere radius $r_{\rm atm}$, we solve the
equations~(\ref{dPdr}), (\ref{dMdr}) and~(\ref{dTdr}) with the
boundary conditions described above to get the structure of the
atmosphere and its mass $M_{\rm atm}$.  We start the integration at
$r=r_{\rm atm}$ with $T=T_s$, $P=P_s$ and an initial guess for
$M(r_{\rm atm})=M_c + M_{\rm atm}$.  In principle, to calculate $P_s$,
we need to know $\rho_s$, which we derive from the pressure.  However,
for the values of $T_s$ we obtain, which are smaller than 1000~K, the
opacity $\kappa_s$ does not depend on $\rho_s$.  We can therefore
self--consistently ignore $\rho_s$ when calculating $\kappa_s$.  The
integration is carried through down to $r=r_c$.  In general, the
condition that $M(r_c)=M_c$ will not be satisfied.  An iteration
procedure is then used to adjust the value of $M(r_{\rm atm})$ until
$M(r_c)=M_c$ to a specified accuracy.  When the core mass is larger
than the critical mass $M_{\rm crit}$, no solution can be found,
i.e. no atmosphere at equilibrium can be supported at the surface of a
core which mass exceeds $M_{\rm crit}$.

The opacity is taken from Bell~\& Lin (1994). This
has contributions from dust grains, molecules, atoms and ions.  It is
written in the form $\kappa=\kappa_i \rho^a T^b$ where $\kappa_i$, $a$
and $b$ vary with temperature. 

We take $\dot{M}_c$ to be in the range $10^{-8}$--$10^{-5}$
M$_{\oplus}$~yr$^{-1}$, as this gives timescales for building up a 10
M$_{\oplus}$ core of $10^6$--$10^9$ years.  These values are
consistent with the calculation of $\dot{M}_c$ by Pollack et
al. (1996) and, more recently, by Movshovitz et al. (2010).

Figure~\ref{fig1} shows the mass of the atmosphere {\em versus} that
of the core for $\dot{M}_c=10^{-5}$, $10^{-6}$, $10^{-7}$ and
$10^{-8}$ M$_{\oplus}$~yr$^{-1}$, and for an atmosphere radius $r_{\rm
atm}$ equal to $5 r_c$, $10 r_c$ or $r_L$.  When $r_{\rm atm}= 5 r_c$,
the photospheric temperature and pressure are $T_s=555.3$~K and
$P_s=3.73$~erg~cm$^{-3}$ for $\dot{M}_c=10^{-5}$
M$_{\oplus}$~yr$^{-1}$, and $T_s=98.7$~K and $P_s=20.42$~erg~cm$^{-3}$
for $\dot{M}_c=10^{-8}$ M$_{\oplus}$~yr$^{-1}$.  The curves
corresponding to $r_{\rm atm}=r_L$ are calculated assuming the
temperature and pressure in the midplane of the nebula to be 33~K and
0.025~erg~cm$^{-3}$, respectively, which correspond to a standard
steady state disc model with $\alpha=10^{-2}$ and gas accretion rate
${\dot M} = 10^{-8}$~M$_{\odot}$~yr$^{-1}$, assuming the protoplanet
is at 5~AU from the central star (see Papaloizou \& Terquem 1999).


For $r_{\rm atm}=10 r_c$, the critical core mass is 36, 24, 16 and
10~M$_{\oplus}$ for $\dot{M}_c=10^{-5}$, $10^{-6}$, $10^{-7}$ and $10^{-8}$
M$_{\oplus}$~yr$^{-1}$, respectively.  For $r_{\rm atm}=5 r_c$, the critical
core mass is 55, 37, 25 and 16~M$_{\oplus}$ for these values of
$\dot{M}_c$.  We note that, if the core is at 5~AU from the central
star, $r_{\rm atm}=10 r_c$ and $r_{\rm atm}=5 r_c$ represent about 1\%
and 0.7\%, respectively, of the Roche lobe radius of the protoplanet
when its core reaches the critical mass.

For comparison, in a standard steady state disc model with
$\alpha=10^{-2}$ and gas accretion rate ${\dot M} =
10^{-8}$~M$_{\odot}$~yr$^{-1}$, at 5~AU from the central star, the
critical core mass of a protoplanet with $r_{\rm atm}=r_L$ is 24, 15,
9 and 5 M$_{\oplus}$ for $\dot{M}_c=10^{-5}$, $10^{-6}$, $10^{-7}$ and
$10^{-8}$ M$_{\oplus}$~yr$^{-1}$, respectively.  These values are
almost independent of the conditions in the nebula, and therefore do
not change when ${\dot M}$ and $\alpha$ are varied.

We see that the critical core mass increases when $r_{\rm atm}$
decreases.  This is because the gas in the atmosphere has less
(negative) gravitational energy when $r_{\rm atm}$ is smaller, i.e. it
is more bound, and therefore, for a fixed core accretion luminosity,
it can be supported at equilibrium up to larger core masses.  Also,
the temperatures at the bottom of the atmosphere tend to be lower for
smaller $r_{\rm atm}$, so that more mass can settle near the surface
of the core, which also helps to increase the critical core mass.
Finally, because the surface of the atmosphere is smaller, radiative
losses are smaller which helps supporting the atmosphere at
equilibrium up to larger core masses.  However, for a given core mass
and core accretion luminosity, the mass of the atmosphere decreases as
$r_{\rm atm}$ decreases.

In all the calculations displayed on figure~\ref{fig1}, $r_L$ is
smaller than the Bondi radius $r_B$ and represents the largest radius
the static atmosphere of the protoplanet can expand to.  If the
midplane temperature of the protoplanetary disc were larger, $r_B$
would become smaller than $r_L$.  This is the case, for instance, for
protoplanets less massive than 16 M$_{\oplus}$ when the midplane
temperature is 140~K, which corresponds to a a disc model with
$\alpha=10^{-2}$ and gas accretion rate ${\dot M} =
10^{-7}$~M$_{\odot}$~yr$^{-1}$, at 5~AU from the central star.

\section{Evolution of a protoplanet accreting from a protoplanetary disc}
\label{sec:evolution}

We are now going to consider the evolution of a protoplanet which 
has an atmosphere that does not fill in its Roche lobe.

\subsection{Accretion onto a core}
\label{sec:accretion}

Numerical simulations of low mass protoplanets ($\sim$ a few
M$_{\oplus}$) embedded in protoplanetary discs and which do not fill
entirely their Roche lobe show that they significantly perturb the
Keplerian motion of the fluid in the vicinity of and within their
Roche lobe.  High resolution two dimensional computations by D'Angelo
et al. (2002) indicate that circumplanetary discs may form within the
Roche lobe of cores more massive than about 5 M$_{\oplus}$.
Subsequent three dimensional simulations by the same authors (D'Angelo
et al. 2003; see also Bate et al. 2003), using an isothermal equation
of state, show that the circumplanetary disc features are much less
marked around low mass cores due to the vertical motion of the fluid
around them.  Gas entering the Roche lobe from above and below the
protoplanet falls down directly onto it instead of joining a
circumplanetary disc in the midplane. However, in these calculations,
the resolution is limited to be several percent of the Hill radius of
the protoplanet, which is between one and two orders of magnitude
larger than the radius of the core itself.  Therefore, from these
simulations, the formation of a circumplanetary disc around low mass
cores cannot be ruled out.  Simulations by Machida et al. (2008) and
Machida (2009) of circumplanetary disc formation around protoplanets
show indeed that these discs tend to be rather compact, with an outer
radius smaller than about 50~core radii.

Further three dimensional numerical simulations by Paardekooper \&
Mellema (2008), including radiative transfer, show significant
differences compared to the isothermal case.  In particular, the
accretion rate onto the protoplanet is much lower due to the
compression of gas within the planetary atmosphere (which, in these
calculations, is the unresolved region around the core).  The authors
comment that the formation of a hot bubble around the protoplanet
limits the gas flow towards its surface.  In these simulations, the
resolution is again on the order of 1\% of the Hill radius, so that
the potential formation of a circumplanetary disc close to the core
could not be seen or dismissed.


We suppose here that the protoplanet accretes through either a cold
Bondi type accretion flow or a circumplanetary disc.  The surface of
the protoplanet atmosphere is at a radius $r_{\rm atm}$ which is the
transition from the (spherical or disc--like) accretion flow to a
hydrostatic model.  The value of the gas accretion rate $\dot{M}_{\rm
gas}$ onto the protoplanet is very uncertain.  The simulations by D'Angelo et
al. (2003) indicate an infall rate onto a 5~M$_{\oplus}$ core on the
order of $10^{-4}$ M$_{\oplus}$~yr$^{-1}$, but computations
incorporating radiative transport give a value which is an order of
magnitude smaller (Paardekooper \& Mellema 2008).  Computations with
higher resolution would give even smaller values as compression of the
gas near the core would be larger on smaller scales.  

When the gas accreted encounters the surface of the protoplanet
atmosphere, it releases the luminosity $L_{\rm acc} = \left[ G \left(
M_c + M_{\rm atm} \right) \dot{M}_{\rm gas} \right] / r_{\rm atm}.$ We
assume that essentially all this luminosity is converted into heat in
and radiated away from the accretion shocks that form at the surface
of the protoplanet (we neglect the energy that is converted into
rotational energy of the protoplanet). The gas that is accreted though
has to bring some entropy deep inside the atmosphere, since the
temperature and pressure rise there as the atmosphere mass increases,
and it does so by inducing some gravitational contraction of the
atmosphere.  This results in an increase of the internal energy of the
atmosphere which in turn drives the expansion of its surface radius.
However, since the mass of gas which is accreted is brought to the
surface of the atmosphere, the resulting increase in the protoplanet
luminosity is not very important (see discussion in section 7.2 of
Papaloizou \& Nelson~2005).  We will neglect it here and will
therefore consider that the accretion of gas does not modify the
luminosity of the protoplanet, which is always given by $L_{\rm
core}$.

In section~\ref{sec:boundary}, we have assumed that the temperature
and pressure at the surface of the static atmosphere were given by the
photospheric values.  This is a valid approximation as long as the
temperature and pressure of the accretion flow are low compared to the
photospheric values for the planet models.
Note that if some of the accretion luminosity were not radiated away,
which may well be the case, the surface temperature $T_s$ would be
higher than the photospheric value.  However, we remark that the
structure of the atmosphere is not very sensitive to $T_s$ for the
masses considered here and, in any case, an increase of $T_s$ would
make the atmosphere more extended and therefore would strengthen the
point we make in this paper.  As far as the surface pressure is
concerned, the only way the infalling gas contributes to it is through
its ram pressure $P_{\rm ram}$, because its motion is supersonic
(e.g., Paardekooper \& Mellema 2008).  When the accretion is
spherically symmetric, the ram pressure is given by:

\begin{equation}
P_{\rm ram} = \frac{\dot{M}_{\rm gas}}{4 \pi r_{\rm atm}^2} \left[ \frac{2G \left(
M_c + M_{\rm atm} \right)}{r_{\rm atm}} \right]^{1/2},
\end{equation}

\noindent where we have assumed that the gas is accelerated to
free--fall velocity by the gravitational potential of the protoplanet.
In the calculations discussed in section~\ref{sec:evolution_num}
below, we have checked that $P_{\rm ram}$ is always smaller (and in
most cases much smaller) than the photospheric pressure $P_s$.

How exactly the gas is accreted by the protoplanet is not known, but
we assume that once the gas is accreted it is redistributed in a
spherically symmetric way at the surface of the atmosphere.  As we
mentionned above, the gas that is accreted induces some gravitational
contraction of the atmosphere.  However, as the accretion rate is low,
we can consider that the atmosphere stays at quasi hydrostatic and
thermal equilibrium as it accretes.  The timescale for establishing
hydrostatic equilibrium is indeed very short.  If the temperature
everywhere in the atmosphere were equal to that in the midplane of the
protoplanetary disc (this gives a lower limit on the real value),
i.e. between 30 and 140~K for a typical disc at a distance of 5~AU
from the central star, then the sound crossing time through a distance
of $10^{12}$~cm, which is a typical value of the Hill radius of a few
earth mass protoplanet, would be on the order of a year.  The amount
of gas accreted on such a timescale is a tiny fraction of the mass of
the atmosphere, and so it can be assumed that it is brought to
equilibrium almost instantaneously.

We now consider the evolution of the protoplanet as it accretes gas at
constant luminosity, assuming quasi hydrostatic and thermal
equilibrium.

\subsection{Evolution of a 5 M$_{\oplus}$ protoplanet}
\label{sec:evolution_num}

For illustrative purposes, we consider a 5 M$_{\oplus}$ core accreting
from a protoplanetary disc.  We calculate a sequence of quasi--static
equilibria along which the mass of the protoplanet increases. As the
gas accretion rate is fixed, constrained by the physics in the disc,
the surface radius $r_{\rm atm}$ of the atmopshere has to adjust to
accomodate the mass increase.  If the mass of the protoplanet
atmosphere, with initially $r_{\rm atm}<r_L$, increases rapidly
relative to that of its core, then, for a constant luminosity, wee see
from figure~\ref{fig1} that $r_{\rm atm}$ has to increase, i.e. the
atmosphere of the protoplanet expands.  In that case, the evolution is
along almost vertical lines going from the lower curve ($r_{\rm atm}=5
r_c$ on these plots) to the upper curve, which corresponds to $r_{\rm
atm}=r_L$. 

For all reasonable values of $\dot{M}_{\rm gas}$ and $\dot{M}_c$, we
have found that a 5 M$_{\oplus}$ core always undergoes such an
evolution and that its atmosphere has to fill in its Roche lobe after a
rather short timescale.  To illustrate this point, we now develop in
detail some particular cases, starting with a core that has $r_{\rm
atm}=5r_c$.

If $\dot{M}_c=10^{-6}$~M$_{\oplus}$~yr$^{-1}$, such a value of $r_{\rm
atm}$ corresponds to $M_{\rm atm}=3.3 \times 10^{-3}$~M$_{\oplus}$. We
assume that the atmosphere accretes at a rate $\dot{M}_{\rm
gas}=10^{-5}$~M$_{\oplus}$~yr$^{-1}$.  After $10^3$ years, the core
mass has hardly changed, but $M_{\rm atm}$ has grown significantly,
being about $1.3 \times 10^{-2}$~M$_{\oplus}$, which corresponds to
$r_{\rm atm}=30r_c$.  After $4 \times 10^3$ years, the atmosphere mass
has grown to $4.3 \times 10^{-2}$~M$_{\oplus}$, which correspond to
$r_{\rm atm}=r_L$.

Let us now consider the case
$\dot{M}_c=10^{-5}$~M$_{\oplus}$~yr$^{-1}$, which gives $M_{\rm
atm}=9.2 \times 10^{-4}$~M$_{\oplus}$ for $r_{\rm atm}=5r_c$, and
$\dot{M}_{\rm gas}=5 \times 10^{-7}$~M$_{\oplus}$~yr$^{-1}$.  After $5 \times
10^3$ years, $M_c=5.05$ M$_{\oplus}$ and $M_{\rm atm}=3.4 \times
10^{-3}$ M$_{\oplus}$, which corresponds to $r_{\rm atm}=25 r_c$.
After $2.5 \times 10^4$ years, $M_c=5.25$ M$_{\oplus}$ and $M_{\rm
atm}=1.3 \times 10^{-2}$ M$_{\oplus}$, which corresponds to $r_{\rm
atm}=r_L$.

Only when $\dot{M}_c$ or $\dot{M}_{\rm gas}$ get very small may we
have models with $r_{\rm atm}<r_L$.  This is the case, for instance,
when $\dot{M}_c=10^{-5}$~M$_{\oplus}$~yr$^{-1}$ and $\dot{M}_{\rm gas}
< 10^{-7}$~M$_{\oplus}$~yr$^{-1}$, or when
$\dot{M}_c=10^{-8}$~M$_{\oplus}$~yr$^{-1}$ and $\dot{M}_{\rm gas} < 5
\times 10^{-5}$~M$_{\oplus}$~yr$^{-1}$.

Therefore, {\em the static atmosphere of a protoplanet of a few earth
masses has to fill in the Roche lobe as gas is being accreted}.  There
is no Bondi type accretion flow around the protoplanet.  The rate at
which gas penetrates from the nebula through the Roche lobe surface in
the atmosphere depends on the rate at which the core grows.  As the
core grows in mass from the accretion of planetesimals, gas from the
nebula is allowed to penetrate in its atmosphere to keep it at
equilibrium.  Unless the core grows very fast or the protoplanet is in
a very low density part of the nebula, there will always be enough gas
coming through the Roche lobe surface to supply the atmosphere so that
it always fills in the Roche lobe.  Only when the core reaches the
critical core mass may the atmosphere contracts so that
$r_{\rm atm}<r_L$. That happens if the nebula cannot deliver gas fast
enough to the collapsing atmosphere.  We therefore conclude that {\em
a protoplanet with a core mass below the critical mass always fills in
the Roche lobe.}

Note that the point we are making in this paper would still be valid
if the luminosity of the protoplanet were not generated by the
accretion of planetesimals but by the contraction of the atmosphere,
provided the Kelvin--Helmholtz timescale were long enough so that the
atmosphere were at quasi--hydrostatic and thermal equilibrium over the
accretion timescale.

We have assumed here that the atmosphere can freely expand up to the
Roche lobe radius.  We now give an argument to show that this is
always the case, even when the Bondi radius is smaller than the Roche
lobe radius.

\subsection{Roche lobe radius {\em vs.} Bondi radius}

It is usually assumed that a protoplanet embedded in a protoplanetary
disc cannot expand beyond the Bondi radius defined as:

\begin{equation}
r_B= \frac{G \left( M_c + M_{\rm atm} \right)}{c^2},
\end{equation}

\noindent where $c$ is the sound speed in the protoplanetary disc
(e.g., Bodenheimer et al. 2000). The argument is that the gas located
at a distance from the protoplanet larger than $r_B$ has a thermal
energy larger than the gravitational energy that would bind it to the
protoplanet, and therefore cannot be accreted by it.  However, if a
molecule of gas located in the protoplanetary disc beyond $r_B$ is
accelerated toward the protoplanet and collides with it, it may become
bound (and therefore be accreted) like any other molecule coming from
within the Roche lobe as described in section~\ref{sec:accretion}.
Accretion of this molecule only requires that it loses at least part of its
(kinetic plus thermal) energy into shocks during
the collision.  If the molecule hits
and settles into a circumplanetary disc before falling onto the planet,
accretion will happen as energy is radiated away from the disc.  When
the molecule becomes bound to the protoplanet, the pressure gradient
in the atmosphere adjusts itself to balance the gravitational
attraction of the protoplanet, i.e.  hydrostatic equilibrium is
maintained.  At fixed luminosity, the protoplanet has to expand to
accommodate this extra mass, and its final surface radius does not have
to be limited by $r_B$.  For a given mass $M_{\rm atm}$ and surface
temperature $T_s$ of the atmosphere, an equilibrium atmosphere can be
constructed with $r_{\rm atm}>r_B$.

This is exactly similar to the process of star formation.  In a
molecular cloud that collapses onto a star, when particles hit the
surface of the star they have a positive energy.  This is because 
they have some thermal
energy in
addition to the kinetic energy due to the acceleration by the
gravitational potential of the forming star.  
Accretion is possible because at least part of this energy is
dissipated into shocks.  The particle that is accreted
may or may not bring some entropy into the stellar envelope, depending
on whether all or only some of the energy is radiated away, and
the star readjusts to equilibrium (Palla \& Stahler 1991, 1992,
Hosokawa et al. 2010).


We conclude that {\em the protoplanet can always expand to fill in its
Roche lobe, even when the Bondi radius is smaller than the Roche lobe
radius.}

\section{Discussion and conclusion}
\label{sec:discussion}

We have shown above that, if a 5 earth mass protoplanet has a static
atmosphere that does not fill in the Roche lobe but is embedded in a
cold accretion flow, it has to expand as it accretes both
planetesimals and gas from the nebula until it does fill in the Roche
lobe.  The evolution occurs on a timescale shorter than the formation
or migration timescales.  The value of 5 earth masses was chosen to
illustrate the point, but the conclusion of course holds for any
smaller protoplanets.  Therefore, we conclude that protoplanets of a
few earth masses have an atmosphere with a static structure all the
way up to the Roche lobe surface, where it matches to the nebula, and
that remains the case until the core reaches the critical mass or the
gas in the nebula gets depleted.

This result holds whether or not the Bondi radius is smaller than the
Roche lobe radius.  Therefore the Bondi radius does not determine the
extent of the static structure around the core, although it gives the
scale on which the density increases in the atmosphere.  For radii
smaller than the Bondi radius, the structure of the atmosphere is almost
independent of the conditions in the nebula, whereas beyond the Bondi
radius the density does not vary much and the structure is determined
by matching to the pressure and temperature in the nebula.

Numerical simulations which show that low mass protoplanets affect the
flow in the surrounding protoplanetary disc inside their Roche lobe
start with artificial initial conditions.  The flow in the Roche lobe
of the protoplanet is not at equilibrium to start with, as the
gravitational attraction of the core is not balanced.  Therefore, as
seen in the simulations, material from the disc rains down on the
protoplanet (D'Angelo et al. 2003; Paardekooper \& Mellema 2008).
When radiative effects are taken into account, this infall is slowed
down as the gas is being compressed around the protoplanet
(Paardekooper \& Mellema 2008, Ayliffe \& Bate 2009).  When the
luminosity generated by this compression is high enough, an almost
spherical atmosphere forms around low--mass protoplanets, ultimately
filling in their Roche lobe (Ayliffe \& Bate 2009).  This is in
agreement with the results presented in this paper, where the
luminosity is not being produced by the gas compression but by the
accretion of planetesimals.  The results reported in the present paper
indicate that numerical simulations of low--mass cores in discs, when
they cannot result in the build--up of a static atmosphere around the
core (like in the isothermal case), should have as initial conditions
a core surrounded by a static atmosphere extending all the way up to
the Roche lobe radius.  In such a case, there will be no cold
accretion flow around the protoplanet.

It has been argued that the orbital migration of low mass cores
depends very significantly on the torque exerted by the gas within the
Roche lobe of the protoplanet (D'Angelo et al. 2003, Crida et
al. 2009).  The results presented above indicate that this is probably
not the case, as the protoplanet itself fills in its Roche lobe.

In this paper, we have considered the evolution of a protoplanet which
accretes planetesimals from the protoplanetary disc.  Such an
accretion may be reduced significantly before the gas in the nebula
has disappeared, with the consequence that the core reaches the
critical mass and the atmosphere has to contract to provide the energy
radiated at its surface.  In that case, its evolution is that
calculated by Papaloizou \& Nelson (2005). 

Finally, we comment that we have not taken into account here the
rotation of the protoplanet onto itself.  Although this would not
affect the main conclusion of our paper, we point out that there may
be some boundary layer at the surface of the protoplanet's atmosphere
due to a finite relative velocity (shear) between the atmopshere and
the surrounding disk material. The size of the boundary layer would be
determined by the disc's viscosity. \\

We thank S. Balbus for discussions and J. Papaloizou for very useful
comments which have greatly improved the original version of this
paper.  C.T. thanks Princeton University, where the research reported
in this paper was done, for hospitality and support.
T.H. acknowledges support from NSF grant AST 0807432 and NASA grant
NNX08AH24G.

\clearpage

\begin{figure}
  \includegraphics[width=14cm]{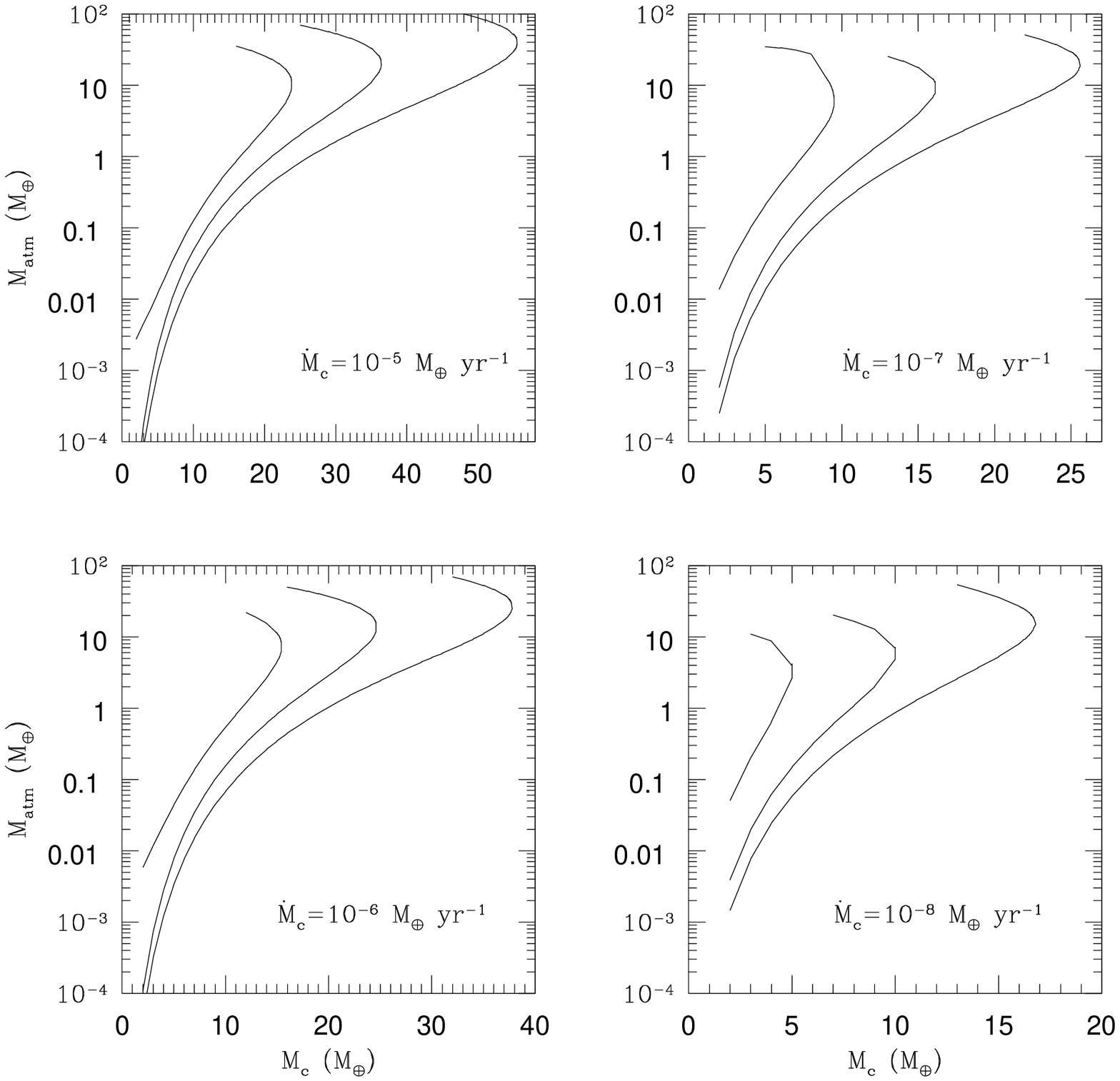} 
  \caption{Atmosphere mass, $M_{\rm atm}$, in units of M$_{\oplus}$,
    vs.  core mass, $M_c$, in units of M$_{\oplus}$, for different
    planetesimal accretion rates $\dot{M}_c$.  The different
    frames correspond to $\dot{M}_c=10^{-5}$ ({\em top left}), 
    $10^{-6}$ ({\em bottom left}), $10^{-7}$ ({\em top right}) and
    $10^{-8}$ ({\em bottom right}) 
    M$_{\oplus}$~yr$^{-1}$, respectively.  Each frame contains three
    curves corresponding, from left to right, to an atmosphere radius
    $r_{\rm atm}=r_L$, $r_{\rm atm}=10 r_c$ 
    and $r_{\rm atm}=5 r_c$,
    respectively. When $r_{\rm atm}=r_L$, the calculation is done for a
    protoplanet embedded in a standard disc model at a distance of
    5~AU from the central star.  The critical core mass is the mass
    beyond which no atmosphere at equilibrium can be supported at the
    surface of the core.  }
     \label{fig1}
\end{figure}

\newpage 

%

\end{document}